\documentstyle[aps,preprint]{revtex}
\tightenlines
\draft
\begin{document}
\draft
\title{Kronecker's Double Series and Exact Asymptotic Expansion for \\
Free Models  of Statistical Mechanics on Torus}
\author{E.V.~Ivashkevich$^{1}$, N.Sh.~Izmailian$^{2,3}$ and
Chin-Kun~Hu$^{2,4}$}
\address{
$^{1}$Bogoliubov Laboratory of Theoretical Physics, J.I.N.R.,
Dubna 141980, Russia\\ $^{2}$Institute of Physics, Academia
Sinica, Nankang, Taipei 11529, Taiwan\\ $^{3}$ Yerevan Physics
Institute, Alikhanian Br. 2, Yerevan 375036, Armenia\\
$^{4}$Department of Physics, National Dong Hwa University, Hualien
97401, Taiwan}
\date{\today}

\maketitle
\begin{abstract}
For the free models of statistical mechanics on torus, exact
asymptotic expansions of the free energy, the internal energy and
the specific heat in the vicinity of the
critical point are found. It is shown that there is direct relation
between the terms of the expansion and the Kronecker's double
series. The latter can be expressed in terms of the elliptic
$\theta$-functions in all orders of the asymptotic expansion.
\end{abstract}
\vspace*{1cm}

\section{Introduction}
\label{introduction}

It is well known that the singularities in thermodynamic functions
associated with a critical point occur only in the thermodynamic
limit when dimension $L$ of the system under consideration tends
to infinity. In such a limit, the critical fluctuations are
correlated over a distance of the order of correlation length
$\xi_{\rm bulk}$ that may be defined as the length scale governing
the exponential decay of correlation functions. Besides these two
fundamental lengths, $L$ and $\xi_{\rm bulk}$, there is also the
microscopic length of interactions $a$. Thermodynamic quantities
thus may in principle depend on the dimensionless ratios $\xi_{\rm
bulk}/L$ and $a/L$. The Finite-Size Scaling (FSS) hypothesis
\cite{FSS} assumes that in the scaling interval, for temperatures
so close to the critical point that $a\ll\xi_{\rm bulk}\sim L$,
the microscopic length drops out and the behaviour of any
thermodynamic quantity can be described in terms of the universal
scaling function of the scaling variable $t=\xi_{\rm bulk}/L$.
However, non-universal corrections to FSS do exist. These
sometimes can be viewed as asymptotic series in powers of $a/L$.

Two-dimensional models of statistical mechanics have long served
as a proving ground in attempts to understand critical behavior
and to test the general ideas of FSS. Very few of them have been
solved exactly \cite{Baxter}. Notably, the so-called free models
that can be treated in terms of non-interacting quasi-particles on
the lattice. The Ising model \cite{McCoyWu} (which is equivalent
to the model of free lattice fermions) and the Gaussian model
(which can be viewed as the model of free lattice bosons) are the
most prominent examples.

For all free models, exact asymptotic expansion of the free energy
on an infinite cylinder of circumference $L$ can easily be
obtained by direct application of the Euler-Maclaurin summation
formula \cite{Cardy,Izmailian}. However, derivation of such an
expansion on a torus of area $S$ and aspect ratio $\rho$ is much
more difficult problem. For the Ising model on torus, such an
expansion has first been studied by Ferdinand and Fisher
\cite{FerdinandFisher}. Starting with the explicit expression for
the partition function \cite{Onsager}, they have calculated two
leading terms, $f_{\rm bulk}$ and $f_0(\rho)$, of the expansion
\begin{equation}
F_{T=Tc}(\rho,S)=f_{\rm bulk}~S + f_0(\rho) +
\sum_{p=1}^{\infty}f_{p}(\rho)~S^{-p} \label{AsymptoticExpansion}
\end{equation}
Their calculations have recently been pushed forward to get two
next sub-leading terms \cite{IzmailianHu,Salas}.

It is the purpose of this paper to derive {\it all} terms of the
exact asymptotic expansion of the logarithm of the partition
function on torus for a class of free exactly solvable models of
statistical mechanics. Our approach is based on an intimate
relation between the terms of the asymptotic expansion and the
so-called Kronecker's double series. Besides the aesthetic appeal
of the exact expansion, there is also physical motivation to study
non-universal corrections to FSS. The problem is that in numerical
simulations of lattice models one usually studies relatively small
lattices. Therefore, to compare the results of high precision
numerical simulations to the theoretical predictions one cannot
neglect sub-leading corrections to FSS \cite{SalasSokal}.
Non-universal terms in the asymptotic expansion also provide
important information about the structure of irrelevant operators
in conformal field theory \cite{Henkel}.

\section{Free Models of Statistical Mechanics}
\label{sec:1}

In this section, we formulate three basic models of statistical
mechanics: Ising model, dimer model and Gaussian model. These
models are often refereed to as "free" since they were shown to be
equivalent to free fermions or free bosons on the lattice.
Partition functions of all these models can be written in terms of
the only object --- the partition function with twisted boundary
conditions $Z_{\alpha,\beta}(\mu)$. Exact asymptotic expansion of
the later will be our main objective in the subsequent section.

\subsection{Ising model}
\label{subsec:1a}

The Ising model is usually formulated as follows. Consider a
planar square lattice of size $M\times N$ with periodic boundary
conditions, i.e. torus. To each site $(m,n)$ of the torus a spin
variable is ascribed, $s_{m\,n}$, with two possible values: $+1$
or $-1$. Two nearest neighbor spins, say $s_{m\,n}$ and
$s_{m\,n+1}$ contribute a term $-J\,s_{m\,n}\,s_{m\,n+1}$ to the
Hamiltonian, where $J$ is some fixed energy. Therefore, the
Hamiltonian is simply the sum of all such terms, one for each edge
of the lattice
\begin{equation}
H(s)=-J\sum_{n=0}^{N-1}\sum_{m=0}^{M-1}
\left(s_{m\,n}\,s_{m+1\,n}+s_{m\,n}\,s_{m\,n+1}\right)
\label{IsingHamiltonian}
\end{equation}
The partition function of the Ising model is given by the sum over
all spin configurations on the lattice
$$ Z_{\rm Ising}(J)=\sum_{\{s\} }e^{-H(s)} $$
It is convenient to set up another parameterizations of the
interaction constant $J$ in terms of the mass variable
$\mu=\ln\sqrt{{\rm sh}\,2J}$. Critical point corresponds to the
massless case $\mu=0$.

An explicit expression for the partition function of the
Ising model on $M \times N$ torus, which was given originally by Kaufmann
\cite{Kaufmann}, can be written as
\begin{equation}
Z_{\rm Ising}(\mu)=\frac{1}{2}\left(\sqrt{2}e^{\mu}\right)^{MN}
\left\{Z_{\frac{1}{2},\frac{1}{2}}(\mu)+
Z_{0,\frac{1}{2}}(\mu)+Z_{\frac{1}{2},0}(\mu)+Z_{0,0}(\mu)\right\}
\label{ZIsing}
\end{equation}
where we have introduced the partition function with twisted
boundary conditions
$$Z^2_{\alpha,\beta}(\mu)=
\prod_{n=0}^{N-1}\prod_{m=0}^{M-1}4\left[\textstyle{
\;\sin^2\left(\frac{\pi (n+\alpha)}{N}\right)+
\sin^2\left(\frac{\pi (m+\beta)}{M}\right)+2\,{\rm sh}^2\mu
\;}\right]$$
Here $\alpha=0$ corresponds to the periodic boundary conditions
for the underlying free fermion in the $N$-direction while
$\alpha=\frac{1}{2}$ stands for anti-periodic boundary conditions.
Similarly $\beta$ controls boundary conditions in $M$-direction.
With the help of the identity \cite{GradshteinRyzhik}
$$ 4\left|~\!{\rm sh}\left(M\omega+i\pi\beta\right)\right|^2
=4\left[\,{\rm sh}^2 M\omega +
\sin^2\pi\beta\,\right]=\prod_{m=0}^{M-1}4\textstyle{
\left[~\!{\rm sh}^2\omega + \sin^2\left(\frac{\pi
(m+\beta)}{M}\right)\right]} $$
the partition function with twisted boundary conditions
$Z_{\alpha,\beta}$ can be transformed into simpler form
\begin{equation}
Z_{\alpha,\beta}(\mu)=\prod_{n=0}^{N-1} 2\left| \textstyle{~\!{\rm
sh}\left[M\omega_\mu\!\left(\frac{\pi(n+\alpha)}{N}\right)+i\pi\beta
\right] }\right| \label{Zab}
\end{equation}
where lattice dispersion relation has appeared
\begin{equation}
\omega_\mu(k)={\rm arcsinh}\sqrt{\sin^2 k+2\,{\rm sh}^2\mu}
\label{SpectralFunction}
\end{equation}
This is nothing but the functional relation between energy
$\omega_\mu$ and momentum $k$ of a free quasi-particle on the
planar square lattice.

\subsection{Dimer model}
\label{subsec:1b}

A "dimer" is a two-atom molecule. When drawn on a lattice it
covers two adjacent sites of the lattice and the bond that joins
them. The "dimer problem" is to determine the number of ways of
covering of a given lattice with dimers, so that all sites are
occupied and no two dimers overlap. If we consider planar square
lattice of size $2M\times 2N$ wrapped on a torus then the number
of dimers must be $2MN$ and the number of distinct dense coverings
of the lattice (the partition function) has been calculated by
Kasteleyn and Fisher \cite{Kasteleyn,Fisher}. This can be
expressed in terms of the same partition function with twisted
boundary conditions Eq.(\ref{Zab}) as
\begin{equation}
Z_{\rm Dimer}=\frac{1}{2}\left\{Z^2_{\frac{1}{2},\frac{1}{2}}(0) +
Z^2_{0,\frac{1}{2}}(0) + Z^2_{\frac{1}{2},0}(0)  -
Z^2_{0,0}(0)\right\} \label{ZDimer}
\end{equation}
First two leading terms of the asymptotic expansion of this
partition function has been obtained by Ferdinand
\cite{Ferdinand}. Let us also mention that dimers are always in
the critical point and have no phase transition.

\subsection{Gaussian model}
\label{subsec:1c}

Let us now turn to a boson analog of Ising model, which is often
referred to as Gaussian model. Again, we consider planar square
lattice of size $N\times M$ wrapped on a torus. To each site
$(m,n)$ of the lattice we assign a continuous variable $x_{m\,n}$.
The Hamiltonian of the model is
\begin{equation}
H(x)=-J\sum_{n=0}^{N-1}\sum_{m=0}^{M-1}
\left(x_{m\,n}\,x_{m+1\,n}+x_{m\,n}\,x_{m\,n+1}-2x^2_{m\,n}\right)
\label{BosonHamiltonian}
\end{equation}
The partition function of the model can be written as
$$Z(J)=\int_{{\bf R}^{MN}} e^{-H(x)}~{\rm d}\sigma(x)$$
If the measure ${\rm d}\sigma(x)$ in the phase space ${{\bf
R}^{MN}}$ is chosen to be Gaussian
$$ {\rm d}\sigma_{\rm
Gauss}(x)=\pi^{-MN/2}\prod_{n=0}^{N-1}\prod_{m=0}^{M-1}
e^{-x^2_{m\,n}}~ {\rm d}x_{m\,n} $$
the integration can be done explicitly and the partition function
of the free boson model can be written in terms of the partition
function with twisted boundary conditions Eq.(\ref{Zab}) and
parameterization $J^{-1}=4\,{\rm ch}^2\,\mu$ as
\begin{equation}
Z_{\rm Gauss}(\mu)=\left(\sqrt{2}\;{\rm
ch}\,\mu\right)^{MN}~\Big[\;Z_{0,0}(\mu)\;\Big]^{-1}
\label{ZBoson}
\end{equation}
This model exhibit phase transition at the point $\mu_c=0$ where
the partition function is divergent. This is due to the presence
of so-called zero mode, i.e. due to the symmetry transformation
$x_{m\,n}\to x_{m\,n}+{\rm const}$, which leave the Hamiltonian
(\ref{BosonHamiltonian}) invariant. Correlation functions of
disorder operator in this model have been studied by Sato, Miwa
and Jimbo \cite{SatoMiwaJimbo}.

The reason why this model is often considered as boson analog of
the Ising model is that, one can choose another measure in the
phase space, which makes this model equivalent to the Ising model
considered above
$$ {\rm d}\sigma_{\rm
Ising}(x)=2^{-MN}\prod_{n=0}^{N-1}\prod_{m=0}^{M-1}
\big[\,\delta(x_{m\,n}-1)+\delta(x_{m\,n}+1)\,\big]~ {\rm
d}x_{m\,n} $$
where $\delta$'s are Dirac $\delta$-functions. With such a
definition the variables $x_{m\,n}$ can actually take only two
values: $+1$ or $-1$, so that $x^2_{m\,n}=1$. In this case
integration can be replaced by summation over discrete values of
$x_{m\,n}=\pm 1$ and the Hamiltonian (\ref{BosonHamiltonian})
coincides with the Hamiltonian of the Ising model
(\ref{IsingHamiltonian}) up to a constant.

\section{Asymptotic Expansion in Terms of Kronecker's Double Series}
\label{sec:2}

In the previous section it was shown that partition functions of
three basic free models of statistical mechanics can be expressed
in terms of the only object: the partition function with twisted
boundary condition $Z_{\alpha,\beta}(\mu)$. In this section we
shall obtain exact asymptotic $1/S$-expansion of the partition
function near the critical point. For reader's convenience, all
the technical details of our calculations and the definitions of
the special functions are summarized in the appendices attached to
the paper.

First of all, let us mention the symmetry properties of the
partition function $Z_{\alpha,\beta}(\mu)$. From its definition
(\ref{Zab}) one can easily verify that it is even and periodic
with respect to its arguments $\alpha$ and $\beta$
\begin{eqnarray}
Z_{\alpha,\beta}(\mu)&=&Z_{\alpha,-\beta}(\mu)
=Z_{-\alpha,\beta}(\mu)\nonumber\\
Z_{\alpha,\beta}(\mu)&=&Z_{1+\alpha,\beta}(\mu)
=Z_{\alpha,1+\beta}(\mu)\nonumber
\end{eqnarray}
These imply that twist angles $\alpha$ and $\beta$ can be taken
from the interval $[0,1]$. Then, one can note that for all twists
$(\alpha,\beta)\neq(0,0)$ the partition function
$Z_{\alpha,\beta}(\mu)$ is even with respect to its mass argument
$\mu$. Hence, near the critical point ($\mu=0$) we have
\begin{equation}
Z_{\alpha,\beta}(\mu)=Z_{\alpha,\beta}(0) +
\frac{\mu^2}{2!}\,Z''_{\alpha,\beta}(0) + \ldots
~~~~~~~(\alpha,\beta)\neq(0,0)
\label{ExpansionZabm}
\end{equation}
The only exception is the point where both $\alpha$ and $\beta$
are equal to zero. This case has to be treated separately since at
this point the partition function turns to zero. As a result, we
have
\begin{equation}
Z_{0,0}(\mu)=\mu\, Z'_{0,0}(0) + \frac{\mu^3}{3!}\,Z'''_{0,0}(0)
+ \ldots ~~~~~~~(\alpha,\beta)=(0,0)
\label{ExpansionZoom}
\end{equation}
In what follows notation $Z_{\alpha,\beta}(\mu)$ will imply
$(\alpha,\beta)\neq(0,0)$.

\subsection{Asymptotic Expansion of $Z_{\alpha,\beta}(0)$}
\label{subsec:2a}

Considering the logarithm of the partition function with twisted
boundary conditions, Eq.(\ref{Zab}), we note, that it can be
transformed as
\begin{equation}
\ln Z_{\alpha,\beta}(0)= M\sum_{n=0}^{N-1}
\omega_0\!\left(\textstyle{\frac{\pi(n+\alpha)}{N}}\right)+
\sum_{n=0}^{N-1}\ln\left|\,1-e^{-2\big[\,M
\omega_0\left(\frac{\pi(n+\alpha)}{N}\right)+i\pi\beta\,\big]}\right|
\label{lnZab}
\end{equation}
The second sum here vanishes in the formal limit $M\to\infty$ when
the torus turns into infinitely long cylinder of circumference
$N$. Therefore, the first sum gives the logarithm of the partition
function with twisted angle $\alpha$ on that cylinder. Its
asymptotic expansion can be found with the help of the
Euler-Maclaurin summation formula (Appendix
\ref{EulerMaclaurinFormula})
\begin{equation}
M\sum_{n=0}^{N-1}\omega_0\!\left(\textstyle{\frac{\pi(n+\alpha)}{N}}\right)=
\frac{S}{\pi}\int_{0}^{\pi}\!\!\omega_0(x)~\!{\rm
d}x-\pi\lambda\rho\,{\rm B}_{2}^\alpha- 2\pi\rho\sum_{p=1}^{\infty}
\left(\frac{\pi^2\rho}{S}\right)^{p}
\frac{\lambda_{2p}}{(2p)!}\;\frac{{\rm B}_{2p+2}^\alpha}{2p+2}
\label{EulerMaclaurinTerm}
\end{equation}
Where $\int_{0}^{\pi}\!\!\omega_0(x)~\!{\rm
d}x = 2 \gamma$, $\gamma =0.915965...$ is Catalan's constant and
${\rm B}^{\alpha}_{p}$ are so-called Bernoulli polynomials.
We have also used the symmetry property,
$\omega_0(k)=\omega_0(\pi-k)$, of the lattice dispersion relation
(\ref{SpectralFunction}) and its Taylor expansion
\begin{equation}
\omega_0(k)=k\left(\lambda+\sum_{p=1}^{\infty}
\frac{\lambda_{2p}}{(2p)!}\;k^{2p}\right)
\label{SpectralFunctionExpansion}
\end{equation}
where $\lambda=1$, $\lambda_2=-2/3$, $\lambda_4=4$, etc. In what
follows, we shall not use the special values of these coefficients
assuming the possibility for generalizations.

The second term in Eq.(\ref{lnZab}) we may transform as
\begin{eqnarray}
&&\sum_{n=0}^{N-1}\ln\left|\,1-e^{-2\big[\,M\omega_0\left(\frac{\pi(n
+\alpha)}{N}\right)+i\pi\beta\,\big]}\right|=\nonumber\\&&~~~~~~~~~~
=-{\tt~Re}\sum_{m=1}^{\infty}\frac{1}{m}\left\{\sum_{n=0}^{[N/2]-1}
e^{-2m\big[\,M\omega_0\left(\frac{\pi(n + \alpha)}{N}\right) +
i\pi\beta\,\big]}\right.\nonumber\\&&~~~~~~~~~~~~~~~~~~~~~~~~~~~~~~~~~~
+\left.\sum_{n=0}^{N-[N/2]-1}
e^{-2m\big[\,M\omega_0\left(\frac{\pi(n + 1 - \alpha)}{N}\right) +
i\pi\beta\,\big]}\right\} \label{Z2exp}
\end{eqnarray}
The argument of the first exponent can be expanded in powers of
$1/S$ if we replace the lattice dispersion relation $\omega_0(x)$
with its Taylor expansion (\ref{SpectralFunctionExpansion})
$$ \exp\left\{-2\pi
m\big[\,\lambda\rho(n+\alpha)+i\beta\,\big]-2\pi m\rho
\sum_{p=1}^{\infty}\frac{\lambda_{2p}}{(2p)!}
\left(\frac{\pi^2\rho}{S}\right)^{p}(n + \alpha)^{2p+1}\right\}$$
Taking into account the relation between moments and cumulants
(Appendix \ref{MomentsCumulants}), we obtain asymptotic expansion
of the first exponent itself in powers of $1/S$
\begin{eqnarray}
&&e^{-2m\big[\,M\omega_0\left(\frac{\pi(n+\alpha)}{N}\right)+i\pi\beta\,\big]}=
e^{-2\pi
m\big[\,\lambda\rho(n+\alpha)+i\beta\,\big]}\nonumber\\[0.15cm]
&&~~~~~~~~~~~~~-2\pi m\rho\sum_{p=1}^{\infty}
\left(\frac{\pi^2\rho}{S}\right)^{p}\frac{\Lambda_{2p}}{(2p)!}
~(n+\alpha)^{2p+1}e^{-2\pi
m\big[\,\lambda\rho(n+\alpha)+i\beta\,\big]}\nonumber
\end{eqnarray}
The differential operators $\Lambda_{2p}$ that have appeared here
can be expressed via coefficients $\lambda_{2p}$ of the expansion
of the lattice dispersion relation as
(\ref{SpectralFunctionExpansion})
\begin{eqnarray}
{\Lambda}_{2}&=&\lambda_2\nonumber\\
{\Lambda}_{4}&=&\lambda_4+3\lambda_2^2\,\frac{\partial}{\partial\lambda}\nonumber\\
{\Lambda}_{6}&=&\lambda_6+15\lambda_4\lambda_2\,\frac{\partial}{\partial\lambda}
+15\lambda_2^3\,\frac{\partial^2}{\partial\lambda^2}\nonumber\\
&\vdots&\nonumber\\ {\Lambda}_{p}&=&\sum_{r=1}^{p}\sum
\left(\frac{\lambda_{p_1}}{p_1!}\right)^{k_1}\ldots
\left(\frac{\lambda_{p_r}}{p_r!}\right)^{k_r}\frac{p!}{k_1!\ldots
k_r!}\;\frac{\partial^k}{\partial\lambda^k}\nonumber
\end{eqnarray}
Here summation is over all positive numbers $\{k_1\ldots k_r\}$
and different positive numbers $\{p_1,\ldots,p_r\}$ such that $p_1
k_1+\ldots+ p_r k_r=p$ and $k=k_1+\ldots+k_r-1$.

The expansion for the second exponent in Eq.(\ref{Z2exp}) can be
obtained along the same lines by substitution: $\alpha\to
1-\alpha$. Plugging the expansion of both of the exponents back
into Eq.(\ref{Z2exp}) we obtain
\begin{eqnarray}
&&\sum_{n=0}^{N-1}\ln\left|\,1-e^{-2\big[\,M\omega_0\left(\frac{\pi(n+\alpha)}{N}\right)
+i\pi\beta\,\big]}\right|=\nonumber\\&&=-\;{\tt
Re}\sum_{m=1}^{\infty}\frac{1}{m}\left\{
\sum_{n=0}^{[N/2]-1}e^{-2\pi m\big[\,\lambda\rho
\left(n+\alpha\right)+i\beta\,\big]}+\!\!\!
\sum_{n=0}^{N-[N/2]-1}\!\!\!e^{-2\pi m\big[\,\lambda\rho
\left(n+1-\alpha\right)+i\beta\,\big]}\right\}\nonumber\\
&&~+~2\pi\rho\sum_{p=1}^{\infty}
\left(\frac{\pi^2\rho}{S}\right)^{p}\frac{\Lambda_{2p}}{(2p)!}\;
{\tt Re}\sum_{m=1}^{\infty}\left\{\sum_{n=0}^{[N/2]-1}
(n+\alpha)^{2p+1}~e^{-2\pi m
\big[\,\lambda\rho(n+\alpha)+i\beta\,\big]}\right.\nonumber\\
&&~~~~~~~~~~~~~~~~~~~~~~~~~~~~~~~+\left.\sum_{n=0}^{N-[N/2]-1}
(n+1-\alpha)^{2p+1}~e^{-2\pi m
\big[\,\lambda\rho(n+1-\alpha)+i\beta\,\big]} \right\}\nonumber
\end{eqnarray}
In all these series, summation over $n$ can be extended to
infinity. The resulting errors are exponentially small and do not
affect our asymptotic expansion in any finite power of $1/S$.

The key point of our analysis is the observation that all the
series that have appeared in such an expansion can be obtained by
resummation of either elliptic theta function,
$\theta_{\alpha,\beta}(\tau)$ (Appendix \ref{ThetaFunctions}), or
Kronecker's double series, ${\rm K}_{p}^{\alpha,\beta}(\tau)$
(Appendix \ref{KroneckerDoubleSeries}). Namely, with the help of
the identities (\ref{IdentityTheta}) and (\ref{IdentityKronecker})
we obtain
\begin{eqnarray}
&&\sum_{n=0}^{N-1}\ln\left|\,1-e^{-2\big[\,M\omega_0\left(\frac{\pi(n+\alpha)}{N}\right)
+i\pi\beta\,\big]}\right|=
\ln\left|\frac{\theta_{\alpha,\beta}(i\lambda\rho)}{\eta(i\lambda\rho)}\right|+\pi\lambda\rho\,{\rm
B}_2^\alpha\nonumber\\[0.1cm]&&~~~~~~~~~~~~~~~~~-2\pi\rho\sum_{p=1}^{\infty}
\left(\frac{\pi^2\rho}{S}\right)^{p}\frac{\Lambda_{2p}}{(2p)!}
\,\frac{{\tt Re}\;{\rm K}_{2p+2}^{\alpha,\beta}(i\lambda\rho)-{\rm
B}_{2p+2}^\alpha}{2p+2} \label{ln(1-e)}
\end{eqnarray}
Substituting Eqs.(\ref{EulerMaclaurinTerm}) and (\ref{ln(1-e)})
into Eq.(\ref{lnZab}) we finally obtain exact asymptotic expansion
of the logarithm of the partition function with twisted boundary
conditions in terms of the Kronecker's double series
\begin{eqnarray}
\ln
Z_{\alpha,\beta}(0)&=&\frac{S}{\pi}\int_{0}^{\pi}\!\!\omega_0(x)~\!{\rm
d}x +
\ln\left|\frac{\theta_{\alpha,\beta}(i\lambda\rho)}{\eta(i\lambda\rho)}\right|\nonumber\\[0.1cm]
&&-2\pi\rho\sum_{p=1}^{\infty}
\left(\frac{\pi^2\rho}{S}\right)^{p}\frac{\Lambda_{2p}}{(2p)!}\,
\frac{{\tt Re}\;{\rm K}_{2p+2}^{\alpha,\beta}(i\lambda\rho)}{2p+2}
\label{ExpansionOflnZab}
\end{eqnarray}
Note, that Bernoulli polynomials ${\rm B}_{p}^{\alpha}$ have
finally dropped out from the asymptotic expansion on torus. This
actually means that Kronecker's double series can be considered as
elliptic generalizations of Bernoulli polynomials.

\subsection{Asymptotic Expansion of $Z'_{0,0}(0)$}
\label{subsec:2b}

As it has already been mentioned, we have to treat the case
$(\alpha,\beta)=(0,0)$ separately. Taking the derivative of
Eq.(\ref{Zab}) with respect to mass variable $\mu$ and then
considering limit $\mu\to 0$ we obtain
$$Z'_{0,0}(0)=2\sqrt{2}M\prod_{n=1}^{N-1} 2\left|
\textstyle{~\!{\rm sh}\left[M\omega_0\!\left(\frac{\pi
n}{N}\right)\right] }\right|$$
Asymptotic expansion of this expression can be found along the
same lines as above. In terms of the Kronecker's double series,
the expansion can be written as
\begin{eqnarray} \ln Z'_{0,0}(0)&=&
\frac{S}{\pi}\int_{0}^{\pi}\!\!\omega_0(x)~\!{\rm
d}x+\frac{1}{2}\ln 8\rho S +
2\ln\left|\,\eta(i\lambda\rho)\,\right|\nonumber\\&&-
2\pi\rho\sum_{p=1}^{\infty}
\left(\frac{\pi^2\rho}{S}\right)^p\frac{\Lambda_{2p}}{(2p)!}\frac{{\tt
~Re}~\!{\rm K}^{0,0}_{2p+2}(i\lambda\rho)}{2p+2}
\label{ExpansionOflnZ00}
\end{eqnarray}

\subsection{Asymptotic Expansion of $Z''_{\alpha,\beta}(0)$}
\label{subsec:2c}

The analysis of the $Z''_{\alpha,\beta}(0)$ is a little more involved.
Taking the second derivative of
Eq.(\ref{Zab}) with respect to mass variable $\mu$ and then
considering limit $\mu\to 0$ we obtain

\begin{eqnarray}
\frac{Z''_{\alpha,\beta}(0)}{Z_{\alpha,\beta}(0)}&=&{\tt
Re}M \sum_{n=0}^{N-1}
\omega''_0\!\left(\textstyle{\frac{\pi(n+\alpha)}{N}}\right) {\rm
cth}\left[M\omega_0\!\left(\textstyle{\frac{\pi(n+\alpha)}{N}}\right)
+i\pi\beta\right]
\label{ExpansionZ''} \\
&=&M \sum_{n=0}^{N-1}\omega''_0\!\left(\textstyle{\frac{\pi(n+\alpha)}{N}}
\right)+2~{\tt Re} M \sum_{n=0}^{N-1}\sum_{m=1}^{\infty}
\omega''_0\!\left(\textstyle{\frac{\pi(n+\alpha)}{N}}\right)
e^{-2m\left[M\omega_0\left(\frac{\pi(n + \alpha)}{N}\right)
+ i\pi\beta\right]}
\nonumber
\end{eqnarray}
where $\omega''_0(x)$ is the second derivative of $\omega_{\mu}(x)$ with
respect to $\mu$ at criticality
\begin{eqnarray}
\omega''_0(x) = \frac{2}{\sin{x}\;\sqrt{1+\sin^2{x}}}
\nonumber
\end{eqnarray}
Using Taylor's theorem, the asymptotic expansion of the  $\omega''_0(x)$  can
be written in the following form
\begin{eqnarray}
\omega''_0(x) = \frac{\kappa}{x}\left[1+\sum_{p=1}^{\infty}\frac
{\kappa_{2p}}{(2p)!}x^{2p}
\right]
\nonumber
\end{eqnarray}
where $\kappa=2$, $\kappa_{2}=-2/3$,  $\kappa_{4}=172/15$, etc. Again, in what
follows, we shall not use the special values of these coefficients
assuming the possibility for generalizations.

The first sum in Eq. (\ref{ExpansionZ''}) we may transform as
\begin{eqnarray}
M \sum_{n=0}^{N-1}\omega''_0\!\left(\textstyle{\frac{\pi(n+\alpha)}{N}}
\right) = M \sum_{n=0}^{N-1} f\!\left(\textstyle{\frac{\pi(n+\alpha)}{N}}
\right) + \frac{\kappa S}{\pi} \sum_{n=0}^{N-1} \left(\frac{1}{n+\alpha}+
\frac{1}{n+1-\alpha}\right)
\label{newsum}
\end{eqnarray}
where we have introduce the function $f(x)=\omega''_0(x)-\kappa/x-
\kappa/(\pi - x)$.
This function and all its derivatives are integrable over the interval
$(0,\pi)$. Thus, for the first term in Eq. (\ref{newsum}) we may use again
the Euler-Maclaurin summation formula (Appendix \ref{EulerMaclaurinFormula}),
and after a little algebra we obtain
\begin{eqnarray}
M \sum_{n=0}^{N-1} f\!\left(\textstyle{\frac{\pi(n+\alpha)}{N}}
\right)&=&\frac{S}{\pi}\int_{0}^{\pi}\!\!f(x)~\!{\rm
d}x - \pi\rho\kappa\sum_{p=1}^{\infty}
\left(\frac{\pi^2\rho}{S}\right)^{p-1}
\frac{\kappa_{2p}{\rm B}_{2p}^\alpha}{p(2p)!}
\nonumber\\
&+&\frac{\kappa S}{\pi}\sum_{p=1}^{\infty}
\frac{{\rm B}_{2p}^\alpha}{p}\frac{1}{N^{2p}}
\label{ExpansionF}
\end{eqnarray}
where $\int_{0}^{\pi}\!\!f(x)~\!{\rm d}x = 2\ln{2}-4\ln{\pi}$.
The second sum in Eq.  (\ref{newsum}) can be written in terms of the
digamma function $\psi(x)$.
\begin{eqnarray}
\sum_{n=0}^{N-1} \left(\frac{1}{n+\alpha}+
\frac{1}{n+1-\alpha}\right)= \left[\psi(N+\alpha)+
\psi(N+1-\alpha)-\psi(\alpha)-\psi(1-\alpha)\right]
\label{psi}
\end{eqnarray}
The asymptotic expansion of the digamma function $\psi(x)$ is given by
(see Appendix \ref{PsiSeries})
\begin{equation}
\psi(N+\alpha)=\ln{N}-\sum_{p=1}^{\infty}(-1)^p
\frac{{\rm B}_{p}^\alpha}{p}\frac{1}{N^p}
\label{ExpansionPsi}
\end{equation}
Using the symmetry properties of the Bernoulli polynomials
${\rm B}_{p}^\alpha$, namely, ${\rm B}_{2p}^{1-\alpha}={\rm B}_{2p}^\alpha$
and ${\rm B}_{2p+1}^{1-\alpha}=-{\rm B}_{2p+1}^\alpha$, the Eq. (\ref{psi})
can be rewritten as
\begin{eqnarray}
\sum_{n=0}^{N-1} \left(\frac{1}{n+\alpha}+
\frac{1}{n+1-\alpha}\right)= 2\ln{N}-
\sum_{p=1}^{\infty}
\frac{{\rm B}_{2p}^\alpha}{p}\frac{1}{N^{2p}}-\psi(\alpha)-
\psi(1-\alpha)
\label{psi1}
\end{eqnarray}
Plugging Eqs. (\ref{ExpansionF}) and (\ref{psi1}) back in Eq. (\ref{newsum})
we have finally obtain
\begin{eqnarray}
M \sum_{n=0}^{N-1}\omega''_0\!\left(\textstyle{\frac{\pi(n+\alpha)}{N}}
\right) &=&\frac{2\kappa S}{\pi}\left[\frac{1}{2\kappa}
\int_{0}^{\pi}\!\!f(x)~\!{\rm d}x+\ln{N}-\frac{\psi(\alpha)+\psi(1-\alpha)}{2}
\right]
\nonumber\\
&-& \pi\rho\kappa\sum_{p=1}^{\infty}
\left(\frac{\pi^2\rho}{S}\right)^{p-1}
\frac{\kappa_{2p}{\rm B}_{2p}^\alpha}{p(2p)!}
\label{newsum1}
\end{eqnarray}

Let us now consider the second sum in Eq. (\ref{ExpansionZ''}). Note that
function $\omega''_0(x)$ can be represent as
\begin{equation}
\omega''_0(x)=\frac{\kappa}{x}\;\exp\left\{
{\sum_{p=1}^{\infty}\frac{\varepsilon_{2p}}{(2p)!}}x^{2p}
\right\}
\label{owega''}
\end{equation}
where coefficients $\varepsilon_{2p}$ and $\kappa_{2p}$ are related to each
other through relation between moments and cumulants
(Appendix \ref{MomentsCumulants}). Following along the same lines as in the
section (3.1), the second sum
in Eq. (\ref{ExpansionZ''}) can be written as

\begin{eqnarray}
&&2~{\tt Re} M \sum_{n=0}^{N-1}\sum_{m=1}^{\infty}
\omega''_0\!\left(\textstyle{\frac{\pi(n+\alpha)}{N}}\right)
e^{-2m\left[M\omega_0\left(\frac{\pi(n + \alpha)}{N}\right)
+ i\pi\beta\right]}=
\nonumber\\
&&=\frac{2\kappa S}{\pi}{\tt Re} \sum_{n=0}^{\infty}
\sum_{m=1}^{\infty}\left\{\frac{1}{n+\alpha} e^{-2\pi m\big[\,\rho
\left(n+\alpha\right)+i\beta\,\big]}+\frac{1}{n+1-\alpha}
e^{-2\pi m\big[\,\rho\left(n+1-\alpha\right)+i\beta\,\big]}\right\}
\nonumber\\
&&+\kappa \pi \rho \Omega_2 {\tt Re} \sum_{n=0}^{\infty}
\sum_{m=1}^{\infty}\left\{(n+\alpha) e^{-2\pi m\big[\,\lambda\rho
\left(n+\alpha\right)+i\beta\,\big]}+(n+1-\alpha)
e^{-2\pi m\big[\,\lambda\rho\left(n+1-\alpha\right)+i\beta\,\big]}\right\}
\nonumber\\
&&-\kappa \pi \rho \sum_{p=2}^{\infty}\frac{\Omega_{2p}}{p(2p)!}
\left(\frac{\pi^2 \rho}{S}\right)^{p-1}{\tt Re}\;
{\rm K}_{2p}^{\alpha,\beta}(i\lambda\rho)+
\kappa \pi \rho \sum_{p=2}^{\infty}\frac{\kappa_{2p}
{\rm B}_{2p}^\alpha}{p(2p)!}
\left(\frac{\pi^2 \rho}{S}\right)^{p-1}\label{ExpansionOmega''1}
\end{eqnarray}
The differential operators $\Omega_{2p}$ that have appeared here
can be expressed via coefficients $\omega_{2p}=\varepsilon_{2p}+\lambda_{2p}
\frac{\partial}{\partial\lambda}$ as
\begin{eqnarray}
{\Omega}_{2}&=&\omega_2\nonumber\\
{\Omega}_{4}&=&\omega_4+3\omega_2^2\,
\nonumber\\
&\vdots&
\nonumber
\end{eqnarray}
Let us introduce the function $R_{\alpha,\beta}(\rho)$
\begin{eqnarray}
R_{\alpha,\beta}(\rho)&=&-\frac{\psi(\alpha)+\psi(1-\alpha)}{2}
\label{Ralphabeta}\\
&+&{\tt Re} \sum_{n=0}^{\infty}
\sum_{m=1}^{\infty}\left\{\frac{1}{n+\alpha} e^{-2\pi m\big[\,\rho
\left(n+\alpha\right)+i\beta\,\big]}+\frac{1}{n+1-\alpha}
e^{-2\pi m\big[\,\rho\left(n+1-\alpha\right)+i\beta\,\big]}\right\}
\nonumber
\end{eqnarray}
which first derivative with respect to $\rho$ is given by (see Appendix
\ref{ThetaFunctions}, Eq. (\ref{IdentityTheta2}))
\begin{eqnarray}
\frac{\partial}{\partial\rho} R_{\alpha,\beta}(\rho) =
-2 \frac{\partial}{\partial\rho}
\ln{\left|\theta_{\alpha,\beta}(i\rho)\right|} +
\frac{1}{2\pi}\left(\frac{\partial}{\partial z}
\ln{\left|\theta_{\alpha,\beta}(i\rho)\right|}\right)^2
\end{eqnarray}
For the cases $(\alpha, \beta) = (0,1/2), (1/2,0), (1/2,1/2)$ the second term
 in above equation is equal to zero, and for $R_{\alpha,\beta}(\rho)$ we
obtain
\begin{eqnarray}
R_{\alpha,\beta}(\rho) =
-2 \ln{\left|\theta_{\alpha,\beta}(i\rho)\right|} + C_E + 2\ln{2}
\end{eqnarray}
where $C_E$ is the Euler constant.

With the help of the identity (\ref{IdentityTheta1}) the Eq.
(\ref{ExpansionOmega''1}) can be rewritten as
\begin{eqnarray}
&&2~{\tt Re} M \sum_{n=0}^{N-1}\sum_{m=1}^{\infty}
\omega''_0\!\left(\textstyle{\frac{\pi(n+\alpha)}{N}}\right)
e^{-2m\left[M\omega_0\left(\frac{\pi(n + \alpha)}{N}\right)
+ i\pi\beta\right]}=
\nonumber\\
&&=\frac{2\kappa S}{\pi}\left[R_{\alpha,\beta}(\rho)
+\frac{\psi(\alpha)+\psi(1-\alpha)}{2}\right] +\frac{\kappa}{2}
\left(\kappa_2\rho\frac{\partial}{\partial\rho}+\lambda_2\rho^2
\frac{\partial^2}{\partial\rho^2}\right)
\ln\left|\frac{\theta_{\alpha,\beta}(i\rho)}{\eta(i\rho)}\right|
\nonumber\\
&&-\kappa \pi \rho \sum_{p=2}^{\infty}\frac{\Omega_{2p}}{p(2p)!}
\left(\frac{\pi^2 \rho}{S}\right)^{p-1}{\tt Re}\;
{\rm K}_{2p}^{\alpha,\beta}(i\lambda\rho)+
\kappa \pi \rho \sum_{p=1}^{\infty}\frac{\kappa_{2p}
\;{\rm B}_{2p}^\alpha}{p(2p)!}
\left(\frac{\pi^2 \rho}{S}\right)^{p-1}\label{ExpansionOmega''2}
\end{eqnarray}
Substituting Eqs. (\ref{newsum1}) and (\ref{ExpansionOmega''2}) into
Eq. (\ref{ExpansionZ''}) we finally obtain exact
asymptotic expansion of the $Z''_{\alpha,\beta}(0)$
\begin{eqnarray}
\frac{Z''_{\alpha,\beta}(0)}{Z_{\alpha,\beta}(0)}
&=&\frac{2\kappa S}{\pi}\left[\frac{1}{2\kappa}
\int_{0}^{\pi}\!\!f(x)~\!{\rm d}x+\ln{N}+R_{\alpha,\beta}(\rho)\right]
\nonumber\\
&+&\frac{\kappa}{2}
\left(\kappa_2\rho\frac{\partial}{\partial\rho}+\lambda_2\rho^2
\frac{\partial^2}{\partial\rho^2}\right)
\ln\left|\frac{\theta_{\alpha,\beta}(i\rho)}{\eta(i\rho)}\right|
\label{ExpansionZ''fin}\\
&-& \kappa \pi \rho \sum_{p=2}^{\infty}\frac{\Omega_{2p}}{p(2p)!}
\left(\frac{\pi^2 \rho}{S}\right)^{p-1}{\tt Re}\;
{\rm K}_{2p}^{\alpha,\beta}(i\lambda\rho)
\nonumber
\end{eqnarray}

\subsection{Asymptotic Expansion of $Z'''_{0,0}(0)$}
\label{subsec:2d}

Let us now consider the case $(\alpha, \beta) = (0, 0)$.
Eq. (\ref{ExpansionZoom}) implies immediately that
\begin{equation}
\lim_{\mu \to 0}\frac{Z''_{0,0}(\mu)}{Z_{0,0}(\mu)}=
\frac{Z'''_{0,0}(0)}{Z'_{0,0}(0)}
\nonumber
\end{equation}
Taking the third derivative of
Eq.(\ref{Zab}) with respect to mass variable $\mu$ and then
considering limit $\mu\to 0$ we obtain
\begin{eqnarray}
\frac{Z'''_{0,0}(0)}{Z'_{0,0}(0)}&=& 2 M^2-1 + 3 M \sum_{n=1}^{N-1}
\omega''_0\!\left(\textstyle{\frac{\pi n}{N}}\right) {\rm
cth}\left[M\omega_0\!\left(\textstyle{\frac{\pi n}{N}}\right)
\right]
\label{ExpansionZ'''}
\end{eqnarray}

Asymptotic expansion of the $Z'''_{0,0}(0)$ can be found along the
same lines as above. In terms of the Kronecker's double series,
the expansion can be written as
\begin{eqnarray}
\frac{Z'''_{0,0}(0)}{Z'_{0,0}(0)}
&=& \frac{6\kappa S}{\pi}\left[\frac{1}{2\kappa}
\int_{0}^{\pi}\!\!f(x)~\!{\rm d}x+\ln{N}+C_E-\ln{\eta(i\rho)}\right]
\nonumber\\
&+&3\kappa
\left(\kappa_2\rho\frac{\partial}{\partial\rho}+\lambda_2\rho^2
\frac{\partial^2}{\partial\rho^2}\right)
\ln{\eta(i\rho)}-1
\nonumber\\
&-& 3\kappa \pi \rho \sum_{p=2}^{\infty}\frac{\Omega_{2p}}{p(2p)!}
\left(\frac{\pi^2 \rho}{S}\right)^{p-1}
{\rm K}_{2p}^{0,0}(i\lambda\rho)
\label{ExpansionZ'''fin}
\end{eqnarray}

Expansions (\ref{ExpansionOflnZab}), (\ref{ExpansionOflnZ00}),
(\ref{ExpansionZ''fin}) and (\ref{ExpansionZ'''fin})
are the main results of the paper. Kronecker's double series ${\rm
K}_{p}^{\alpha,\beta}$ with $\alpha$ and $\beta$ taking values $0$
and $1/2$ can all be expressed in terms of the elliptic
$\theta$-functions only (Appendix~\ref{KroneckerToTheta}).

\section{Asymptotic Expansion of the Free Energy,\\
the Internal Energy and the Specific Heat}
\label{subsec:2c}

After reaching this point, one can easily write down all the terms
of the exact asymptotic expansion (\ref{AsymptoticExpansion}) of
the free energy, $F=-\ln Z$, at the critical point for all three
models under consideration. For the Ising model and the Gaussian model we
have found that the exact asymptotic expansion of the internal energy,
$U = - \partial \ln Z(J)/\partial J$, and
the specific heat, $C = \partial^2 \ln Z(J)/\partial J^2$,
at the critical point can be written in the following form
\begin{eqnarray}
U &=& u_{\rm bulk}~S +
\sum_{p=0}^{\infty} u_p(\rho)~S^{-p+\frac{1}{2}}
\nonumber\\
C &=& c_{\rm bulk}~S + \sum_{p=0}^{\infty} c_{2 p}(\rho)~S^{-p+\frac{1}{2}} +
\sum_{p=0}^{\infty} c_{2 p+1}(\rho)~S^{-p}
\nonumber
\end{eqnarray}

\begin{enumerate}
\item
The asymptotic expansion of the free energy of the Ising model is

\begin{eqnarray}
f_{\rm bulk}&=& -\ln\sqrt{2}-\frac{2\gamma}{\pi}
\nonumber\\
f_0(\rho)&=& -\ln\frac{\theta_2+\theta_3+\theta_4}{2\eta}
\nonumber \\
f_1(\rho)&=&-\frac{\pi^3
\rho^2}{180}~\frac{\frac{7}{8}({\theta}_2^9+{\theta}_3^9+{\theta}_4^9)+
{\theta}_2 {\theta}_3 {\theta}_4 \left[ {\theta}_2^3
{\theta}_4^3-{\theta}_3^3{\theta}_2^3-{\theta}_3^3{\theta}_4^3
\right]}{{\theta}_2+{\theta}_3+{\theta}_4}
\nonumber \\
f_2(\rho)&=& -\frac{\pi^6 \rho^5}{18432}~
\frac{{\theta}_2 {\theta}_3 {\theta}_4\left\{{\theta}_3^7({\theta}_2^4-{\theta}_4^4)^2
+{\theta}_2^7({\theta}_3^4+{\theta}_4^4)^2+{\theta}_4^7({\theta}_3^4+
{\theta}_2^4)^2\right\}}{({\theta}_2+{\theta}_3+{\theta}_4)^2}
\nonumber\\
&-&\frac{31\pi^5 \rho^4}{24192}~
\frac{{\theta}_3^4({\theta}_4^9-{\theta}_2^9)
+{\theta}_2^4({\theta}_4^9-{\theta}_3^9)+{\theta}_4^4({\theta}_3^9-
{\theta}_2^9)}{{\theta}_2+{\theta}_3+{\theta}_4}
\left(1+4\rho \frac{\partial}{\partial \rho}
\ln{\theta_2}\right)
\nonumber\\
&-&\frac{\pi^5 \rho^4}{1512}~
\frac{{\theta}_2 {\theta}_3 {\theta}_4
\left[{\theta}_3^3({\theta}_2^7-{\theta}_4^7)
+{\theta}_2^3({\theta}_3^7-{\theta}_4^7)+{\theta}_4^3({\theta}_2^7-
{\theta}_3^7)\right]
}
{{\theta}_2+{\theta}_3+{\theta}_4}
\left(1+4\rho \frac{\partial}{\partial \rho}
\ln{\theta_2}\right)
\nonumber \\
&-& \frac{31\pi^6 \rho^5}
{72576}~
\frac{{\theta}_3^4{\theta}_4^4\left[
3{\theta}_3^4{\theta}_4^4
({\theta}_3+{\theta}_4)-{\theta}_2^4(2{\theta}_2^5+
{\theta}_3^5 -{\theta}_4^5)\right]}{{\theta}_2+{\theta}_3+{\theta}_4}
\nonumber\\
&-& \frac{\pi^6 \rho^5}
{4536}~
\frac{{\theta}_3^4{\theta}_4^4\left[{\theta}_2 {\theta}_3
({\theta}_2^7 -{\theta}_3^7)+{\theta}_2 {\theta}_4
({\theta}_2^7 -{\theta}_4^7)
+2{\theta}_2^4 {\theta}_3{\theta}_4
({\theta}_3^3-{\theta}_4^3)-4{\theta}_2 {\theta}_3^4{\theta}_4^4\right]
}{{\theta}_2+{\theta}_3+{\theta}_4}
\nonumber\\
&\vdots&
\nonumber
\end{eqnarray}

\begin{eqnarray}
u_{\rm bulk}&=& -\sqrt{2}
\nonumber\\
u_0(\rho)&=& -2\sqrt{\rho}~
\frac{{\theta}_2{\theta}_3{\theta}_4}{{\theta}_2+{\theta}_3+{\theta}_4}
\nonumber\\
u_1(\rho)&=&\frac{\pi^3 \rho^{5/2}}{48}~ \frac{
{\theta}_2{\theta}_3{\theta}_4({\theta}_2^9+{\theta}_3^9+{\theta}_4^9)}{
({\theta}_2+{\theta}_3+{\theta}_4)^2}
\nonumber\\
u_2(\rho)&=&\frac{\pi^6\rho^{9/2}}{4608}~
\frac{{\theta}_2^2{\theta}_3^2{\theta}_4^2
\left[
{\theta}_3^7({\theta}_2^4-{\theta}_4^4)^2+
{\theta}_2^7({\theta}_3^4+{\theta}_4^4)^2+
{\theta}_4^7({\theta}_3^4+{\theta}_2^4)^2
\right]}
{({\theta}_2+{\theta}_3+{\theta}_4)^3}
\nonumber\\
&+&\frac{\pi^6\rho^{9/2}}{9216}~
\frac{{\theta}_2{\theta}_3{\theta}_4
\left[23({\theta}_2^{17}+
{\theta}_3^{17}+{\theta}_4^{17})+8{\theta}_2^4{\theta}_4^4(5{\theta}_2^9+
5{\theta}_4^9-8{\theta}_3^9+2{\theta}_2^4{\theta}_4^5+2{\theta}_2^5
{\theta}_4^4)\right]
}
{({\theta}_2+{\theta}_3+{\theta}_4)^2}
\nonumber\\
&-&\frac{\pi^5\rho^{7/2}}{192}~
\frac{
{\theta}_2{\theta}_3{\theta}_4\left[
{\theta}_3^9({\theta}_2^4-{\theta}_4^4)+
{\theta}_2^9({\theta}_3^4+{\theta}_4^4)-
{\theta}_4^9({\theta}_3^4+{\theta}_2^4)\right]}
{({\theta}_2+{\theta}_3+{\theta}_4)^2}
\left(
1+\frac{\pi\rho}{2}{\theta}_3^4+4\rho
\frac{\partial}{\partial \rho}\ln{\theta_2}\right)
\nonumber\\
&\vdots&
\nonumber
\end{eqnarray}

\begin{eqnarray}
c_{\rm bulk}(\rho)&=&\frac{8}{\pi}
\left(\ln{\sqrt{\frac{S}{\rho}}}+
\ln
{\frac{2^{5/2}}{\pi}}+C_E- \frac{\pi}{4}\right)-4 \rho \left(\frac{
{\theta}_2{\theta}_3{\theta}_4}{{\theta}_2+{\theta}_3+{\theta}_4}\right)^2
-\frac{16}{\pi}\frac{\sum_{i=2}^4{\theta}_i \ln{{\theta}_i}}
{{\theta}_2+{\theta}_3+{\theta}_4}
\nonumber\\
c_0(\rho)&=& -2\sqrt{2}\sqrt{\rho}~\frac{
{\theta}_2{\theta}_3{\theta}_4}{{\theta}_2+{\theta}_3+{\theta}_4}
\nonumber\\
c_1(\rho)&=&\frac{\pi^2 \rho^2}{6}~
\frac{({\theta}_2^8-{\theta}_3^8)
{\theta}_2 {\theta}_3\ln{\frac{{\theta}_3}{{\theta}_2}}+
({\theta}_4^8-{\theta}_3^8)
{\theta}_4 {\theta}_3\ln{\frac{{\theta}_3}{{\theta}_4}}+
({\theta}_2^8-{\theta}_4^8)
{\theta}_4 {\theta}_2\ln{\frac{{\theta}_4}{{\theta}_2}}}
{({\theta}_2+{\theta}_3+{\theta}_4)^2}
\nonumber\\
&+&\frac{\pi^2 \rho^2}{9}~
\frac{{\theta}_3^4 {\theta}_4^4
(2{\theta}_2 -{\theta}_3-{\theta}_4)}{{\theta}_2+{\theta}_3+{\theta}_4}
+\frac{\pi^3 \rho^3}{12}
\frac{{\theta}_2^2 {\theta}_3^2 {\theta}_4^2
({\theta}_2^9+{\theta}_3^9+{\theta}_4^9)}
{({\theta}_2+{\theta}_3+{\theta}_4)^3}
\nonumber\\
&+&\frac{\pi \rho}{9}~
\frac{{\theta}_2^5-{\theta}_4^5+
{\theta}_3({\theta}_2^4-{\theta}_4^4)-2{\theta}_2{\theta}_4
({\theta}_2^3-{\theta}_4^3)}{{\theta}_2+{\theta}_3+{\theta}_4}
\left(1+4\rho\frac{\partial}{\partial \rho}
\ln{\theta_2}\right)
\nonumber\\
c_2(\rho)&=&\frac{\pi^3\rho^{5/2}}{24\sqrt{2}}~
\frac{{\theta}_2 {\theta}_3 {\theta}_4
({\theta}_2^9+{\theta}_3^9+{\theta}_4^9)}
{({\theta}_2+{\theta}_3+{\theta}_4)^2}
\nonumber\\
&\vdots&
\nonumber
\end{eqnarray}
We have also used the following relations between derivatives of the elliptic
functions
$$
\frac{\partial}{\partial \rho}\ln{{\theta}_3} =
\frac{\pi}{4}{\theta}_4^4+\frac{\partial}{\partial \rho}\ln{{\theta}_2}
\qquad \mbox{and} \qquad
\frac{\partial}{\partial \rho}\ln{{\theta}_4} =
\frac{\pi}{4}{\theta}_3^4+\frac{\partial}{\partial \rho}\ln{{\theta}_2}
$$
Note, that with the help of the identities
$$
\frac{\partial}{\partial \rho}\ln{\theta_2} = -\frac{1}{2}{\theta}_3^2 E,
\qquad \mbox{and} \qquad \frac{\partial E}{\partial \rho}
=\frac{\pi^2}{4}{\theta}_3^2{\theta}_4^4-\frac{\pi}{2}{\theta}_4^4 E
$$
one can express all derivatives of the elliptic functions in terms of the
elliptic functions ${\theta}_2, {\theta}_3, {\theta}_4$ and the elliptic
integral of the second kind $E$

\item
Similar expansion of the free energy of the dimer model is
\begin{eqnarray}
f_{\rm bulk}&=& -\frac{\gamma}{\pi}
\nonumber\\
f_0(\rho)&=&-\ln\frac{\theta_2^2+\theta_3^2+\theta_4^2}{2\eta^2}
\nonumber\\
f_1(\rho)&=&-\frac{\pi^3
\rho^2}{90}~
\frac{\frac{7}{8}({\theta}_2^{10}+{\theta}_3^{10}+{\theta}_4^{10})+
{\theta}_2^2 {\theta}_3^2 {\theta}_4^2 \left[
{\theta}_2^2{\theta}_4^2-{\theta}_2^2{\theta}_3^2-{\theta}_3^2{\theta}_4^2
\right]}{{\theta}_2^2+{\theta}_3^2+{\theta}_4^2}
\nonumber\\
f_2(\rho)&=&-\frac{\pi^6 \rho^4}{16200}~
\frac{
{\theta_3^2}\left(\frac{7}{8}{\theta}_3^{8}+
{\theta}_2^{4}{\theta}_4^{4}\right)^2+
{\theta_2^2}\left(\frac{7}{8}{\theta}_2^{8}-
{\theta}_3^{4}{\theta}_4^{4}\right)^2+
{\theta_4^2}\left(\frac{7}{8}{\theta}_4^{8}-
{\theta}_3^{4}{\theta}_2^{4}\right)^2
}
{{\theta}_2^2+{\theta}_3^2+{\theta}_4^2}
\nonumber\\
&+& \frac{\pi^6
\rho^4}{16200}~ \left[\frac{\frac{7}{8}({\theta}_2^{10}+{\theta}_3^{10}+{\theta}_4^{10})+
{\theta}_2^2 {\theta}_3^2 {\theta}_4^2 \left[
{\theta}_2^2{\theta}_4^2-{\theta}_2^2{\theta}_3^2-{\theta}_3^2{\theta}_4^2
\right]}{{\theta}_2^2+{\theta}_3^2+{\theta}_4^2}\right]^2
\nonumber\\
&+&\frac{\pi^6
\rho^4}{756}~\frac{{\theta}_3^{8}({\theta}_2^2-
{\theta}_4^2)+{\theta}_4^{8}({\theta}_2^2-
{\theta}_3^2)+\frac{5}{8}\left[
{\theta}_3^{2}({\theta}_2^8-
{\theta}_4^8)+{\theta}_4^{2}({\theta}_2^8-
{\theta}_3^8)
\right]+\frac{5}{16}(2{\theta}_2^{10}-{\theta}_3^{10}-
{\theta}_4^{10})
}
{{\theta}_2^2+{\theta}_3^2+{\theta}_4^2}
\nonumber\\
&+& \frac{31\pi^5
\rho^3}{12096}~\frac{{\theta}_3^{10}({\theta}_2^4-
{\theta}_4^4)+{\theta}_2^{10}({\theta}_3^4+
{\theta}_4^4)-{\theta}_4^{10}({\theta}_3^4+
{\theta}_2^4)}
{{\theta}_2^2+{\theta}_3^2+{\theta}_4^2}
\left(1+4\rho \frac{\partial}{\partial \rho}
\ln{\theta_2}\right)
\nonumber\\
&-& \frac{\pi^5
\rho^3}{756}~\frac{
{\theta}_2^2{\theta}_3^2{\theta}_4^2
\left[{\theta}_3^{6}({\theta}_2^2-
{\theta}_4^2)+{\theta}_2^{6}({\theta}_3^2+
{\theta}_4^2)-{\theta}_4^{6}({\theta}_3^2+
{\theta}_2^2)\right]}
{{\theta}_2^2+{\theta}_3^2+{\theta}_4^2}
\left(1+4\rho \frac{\partial}{\partial \rho}
\ln{\theta_2}\right)
\nonumber\\
&\vdots&
\nonumber
\end{eqnarray}
\item
Finally, the asymptotic expansion of the free energy, the internal energy
and the specific heat of the Gaussian model after subtraction of the zero
modes, $\ln\mu\sqrt{8S}$, $\sqrt{2}S/\mu$ and $S(1/\mu+1/\mu^2)$,
respectively, can be written as
\begin{eqnarray}
f_{\rm bulk}&=& \frac{2\gamma}{\pi} - \ln\sqrt{2}
\nonumber\\
f_0(\rho)&=&\ln\sqrt{\rho}\,\eta^2
\nonumber\\
f_1(\rho)&=&\frac{\pi^3\rho^2}{180}~
(\theta_2^4\theta_4^4-\theta_2^4\theta_3^4-\theta_3^4\theta_4^4)
\nonumber\\
f_2(\rho)&=& \frac{\pi^6 \rho^4}{1512}~\theta_3^4\theta_4^4 \left
(\theta_2^8-2\theta_3^4\theta_4^4\right)+ \frac{\pi^5 \rho^3}{1512}
(\theta_3^4+\theta_4^4)(\theta_3^4+\theta_2^4)(\theta_2^4-\theta_4^4)
\left(1+4\rho\frac{\partial}
{\partial \rho}\ln{\theta_2}\right)
\nonumber\\
&\vdots&
\nonumber
\end{eqnarray}
For internal energy there are no finite-size correction terms
($u_{bulk}=0, u_p=0$ for all $p$).
\begin{eqnarray}
c_{\rm bulk}(\rho)&=&\frac{8}{\pi}
\left(\ln{\sqrt{\frac{S}{\rho}}}+
\ln
{\frac{\sqrt{2}}{\pi}}+C_E- \frac{\pi}{4}+\ln{\eta}\right)
\nonumber\\
c_0(\rho)&=&0
\nonumber\\
c_1(\rho)&=&\frac{2}{3}\left(1+4\rho \frac{\partial}{\partial \rho}\ln{\eta}
+4\rho^2 \frac{\partial^2}{\partial \rho^2}\ln{\eta}\right)
\nonumber\\
c_2(\rho)&=&0
\nonumber\\
&\vdots&
\nonumber
\end{eqnarray}

\end{enumerate}

\section{Summary}
\label{sec:4}

In this paper, we have derived exact asymptotic expansion of the
partition function with twisted boundary conditions at the
critical point, Eqs. (\ref{ExpansionOflnZab}) and
(\ref{ExpansionOflnZ00}). As an application of this result, we
have obtained exact asymptotic expansion of the free energy, the
internal energy and the specific heat for a
class of free exactly solvable models of statistical mechanics on
torus. Moreover, the partition function of the dimer model on the
Klein bottle can also be expressed via the partition function with
twisted boundary conditions, namely
$Z_{\rm Klein}=Z_{\frac{1}{4},\frac{1}{2}}(0)$
\cite{Klein}. Exact asymptotic expansion of the latter can
immediately be written down with the help of our general formula
(\ref{ExpansionOflnZab}). An interesting question is: whether this
is also the case for other free models? This, however, is the
problem for the future.

\appendix

\section{Euler-Maclaurin Summation Formula.}
\label{EulerMaclaurinFormula}

Suppose that $F(x)$ together with its first $2m$ derivatives is continuos
within the interval $(a, b)$. Then the general Euler-Maclaurin summation
formula states \cite{Abramowitz}

\begin{equation}
\sum_{n=0}^{N-1} F(a+n h+\alpha h)=\frac{1}{h}\int_{a}^{b}
F(\tau)~{\rm d}\tau +
\sum_{k=1}^{p} \frac{h^{k-1}}{k!}
{\rm B}_{k}(\alpha)\left(F^{(k-1)}(b)-F^{(k-1)}(a)\right)-{\rm R}_p(\alpha)
\label{EMFormula}
\end{equation}
where $p\le 2 m$, $0 \le \alpha \le 1$, $h=(b-a)/N$ and reminder term
${\rm R}_p(\alpha)$ is given by
\begin{equation}
{\rm R}_p(\alpha)=\frac{h^p}{p!}\int_{0}^{1}\hat{{\rm B}}_{p}(\alpha-\tau)
\left\{\sum\limits_{n=0}^{N-1}F^{(p)}(a+n h+\tau h)\right\}{\rm d}\tau
\label{Reminder}
\end{equation}
$\hat{{\rm B}}_{p}(\alpha)$ are so called periodic Bernoulli functions
which are defined as follows
\begin{equation}
\hat{{\rm B}}_p(\alpha)=-\frac{p!}{(-2\pi i)^p}\sum_{n\neq
0}\frac{e^{-2\pi in \alpha}}{n^{p}} \label{FourierBernoulli}
\end{equation}
These functions have singularities at the integer values of $\alpha$ and
on the interval $\alpha\in[0,1]$ they coincide with so-called Bernoulli
polynomials
\begin{eqnarray}
{\rm B}_{1}(\alpha) &=& \alpha-\textstyle{\frac{1}{2}}\nonumber\\
{\rm B}_{2}(\alpha) &=&
\alpha^2-\alpha+\textstyle{\frac{1}{6}}\nonumber\\
{\rm B}_{3}(\alpha) &=&
\alpha^3-\textstyle{\frac{3}{2}}\alpha^2+\textstyle{\frac{1}{2}}\alpha
\nonumber\\
{\rm B}_{4}(\alpha) &=&
\alpha^4-2\alpha^{3}+\alpha^2-\textstyle{\frac{1}{30}}\nonumber
\end{eqnarray}
The generating function for the Bernoulli polynomials is
$$\frac{\lambda\,e^{\lambda\alpha}}{e^{\lambda}-1}=
1+\sum_{p=1}^{\infty}\frac{\lambda^{p}}{p!}\,{\rm B}_p(\alpha)$$
Fourier transform of the generating function gives the important
identity
\begin{equation}
\frac{e^{2\pi iz\alpha}}{e^{2\pi
iz}-1}=-\sum_{n=0}^{\infty}e^{2\pi iz(n+\alpha)}=\frac{1}{2\pi i}
\sum_{n=-\infty}^{+\infty}\frac{e^{-2\pi i n \alpha}}{z + n}
\label{FourierGeneratingFunction}
\end{equation}
It is well known that Euler-Maclaurin summation formula
is closely related to (generally speaking divergent) asymptotic series.
For further discussion of the properties of these series the interested
reader is referred to the book of Hardy \cite{Hardy}.

In this paper we are mainly interested in sums of the form

\begin{equation}
\frac{1}{N}\sum_{n=0}^{N-1} f\left(\frac{\pi (n+\alpha)}{N}\right)
\label{EM1Formula}
\end{equation}
The asymptotic expansion of the sum (\ref{EM1Formula}) in the limit
$N\to\infty$ can be obtained from Eq. (\ref{EMFormula}) by setting
$(a=0, b=\pi)$. If we assume that all the derivatives of $f(x)$ are
integrable over the interval $(0, \pi)$, i.e., the integral in Eq.
(\ref{Reminder}) is finite, we can formally extend the sum in
Eq. (\ref{EMFormula}) to $k=\infty$ and drop the reminder term
${\rm R}_p(\alpha)$. In this case
we can write the asymptotic expansion of the sum (\ref{EM1Formula}) as
follows
\begin{equation}
\frac{1}{N}\sum_{n=0}^{N-1} f\left(\frac{\pi (n+\alpha)}{N}\right)=
\frac{1}{\pi}\int_{0}^{\pi}
f(\tau)~{\rm d}\tau +
\frac{1}{\pi}\sum_{k=1}^{\infty} \left(\frac{\pi}{N}\right)^{k}\frac{{\rm B}_{k}(\alpha)}{k!}
\left(f^{(k-1)}(\pi)-f^{(k-1)}(0)\right)
\label{EM2Formula}
\end{equation}

Note, that we also use abbreviated notation for Bernoulli polynomials:
${\rm B}_{p}(\alpha)={\rm B}_{p}^{\alpha}$.

\section{Relation between moments and cumulants}
\label{MomentsCumulants}

Moments $Z_k$ and cumulants $F_k$ which enters the expansion of
exponent
$$\exp\left\{\;\sum_{k=1}^{\infty}\frac{x^k}{k!}\,F_k\,\right\}
=1+\sum_{k=1}^{\infty}\frac{x^k}{k!}\,Z_k$$
are related to each other as \cite{Prohorov}
\begin{eqnarray}
Z_1&=&F_1\nonumber\\ Z_2&=&F_2+F^2_1\nonumber\\
Z_3&=&F_3+3F_1F_2+F^3_1\nonumber\\
Z_4&=&F_4+4F_1F_3+3F_2^2+6F_1^2F_2+F^4_1\nonumber\\
~&\vdots&~\nonumber\\ Z_k&=&\sum_{r=1}^{k}\sum
\left(\frac{F_{k_1}}{k_1!}\right)^{i_1}\ldots
\left(\frac{F_{k_r}}{k_r!}\right)^{i_r} \frac{k!}{i_1!\ldots
i_r!}\nonumber
\end{eqnarray}
where summation is over all positive numbers $\{i_1\ldots i_r\}$
and different positive numbers $\{k_1,\ldots,k_r\}$ such that $k_1
i_1+\ldots+ k_r i_r=k$.

\section{Elliptic Theta Functions}
\label{ThetaFunctions}

We adopt the following definition of the elliptic
$\theta$-functions:
\begin{eqnarray}
\theta_{\alpha,\beta}(z,\tau)&=&\sum_{n\in Z} \exp\left\{ \pi
i\tau \left(n+\textstyle{\frac{1}{2}}-\alpha\right)^2+2\pi i
\left(n+\textstyle{\frac{1}{2}}-\alpha\right)\left(z+\textstyle{\frac{1}{2}}-\beta\right)
\right\}~~~~\nonumber\\ &=&\eta(\tau)\,\exp\left\{\textstyle{\pi
i\tau\big(\alpha^2-\alpha+\frac{1}{6}\big)+2\pi
i\big(\frac{1}{2}-\alpha\big)\big(z+\frac{1}{2}-\beta\big)}\right\}\nonumber\\
&\times&\prod_{n=0}^{\infty}\!\Big[\,1-e^{2\pi
i\tau\left(n+\alpha\right)-2\pi i\left(z-\beta\right)}\,\Big]
\Big[\,1-e^{2\pi i\tau\left(n+1-\alpha\right)+2\pi
i\left(z-\beta\right)}\,\Big]\nonumber
\end{eqnarray}
These should be compared with the notations of Mumford
\cite{Mumford}.

The elliptic $\theta$-functions satisfies the heat equation
\begin{equation}
\frac{\partial}{\partial \tau}\theta_{\alpha,\beta}(z,\tau) =
\frac{1}{4\pi i}\frac{\partial^2}{\partial z^2}\theta_{\alpha,\beta}(z,\tau)
\label{heat}
\end{equation}
Relation to standard notations is
\begin{eqnarray}
\theta_{0,0}(z,\tau)&=&\theta_1(z,\tau)\nonumber\\
\theta_{0,\frac{1}{2}}(z,\tau)&=&\theta_2(z,\tau)\nonumber\\
\theta_{\frac{1}{2},0}(z,\tau)&=&\theta_4(z,\tau)\nonumber\\
\theta_{\frac{1}{2},\frac{1}{2}}(z,\tau)&=&\theta_3(z,\tau)\nonumber
\end{eqnarray}
The Dedekind $\eta$-function is usually defined as
$$\eta(\tau)=e^{\pi i\tau/12}\prod_{n=1}^{\infty}\Big[\,1-e^{2\pi
i \tau n}\,\Big]$$ Considering the functions
$\theta_{\alpha,\beta}(\tau)=\theta_{\alpha,\beta}(0,\tau)$ and
$\eta(\tau)$ of pure imaginary aspect ratio, $\tau=i\rho$, we
obtain the identity
\begin{eqnarray}
\ln\left|\frac{\theta_{\alpha,\beta}(i\rho)}{\eta(i\rho)}\right|+\pi\rho\,{\rm
B}^\alpha_2&=&\sum_{n=0}^{\infty}\ln\left|\,1-e^{-2\pi\rho
\left(n+\alpha\right)-2\pi i\beta}\,\right|\nonumber\\
&&+\sum_{n=0}^{\infty}\ln\left|\,1-e^{-2\pi\rho
\left(n+1-\alpha\right)-2\pi i\beta}\,\right|
\label{IdentityTheta}
\end{eqnarray}
Taking the derivative of Eq. (\ref{IdentityTheta}) with respect to $\rho$
we can obtain the following useful identity
\begin{eqnarray}
&&{\tt Re} \sum_{n=0}^{\infty}
\sum_{m=1}^{\infty}\left[(n+\alpha) e^{-2\pi m\big[\,\rho
\left(n+\alpha\right)+i\beta\,\big]}+(n+1-\alpha)
e^{-2\pi m\big[\,\rho\left(n+1-\alpha\right)+i\beta\,\big]}\right]
\nonumber\\
&&=
\frac{{\rm
B}^\alpha_2}{2}+\frac{1}{2\pi}\frac{\partial}{\partial\rho}
\ln\left|\frac{\theta_{\alpha,\beta}(i\rho)}{\eta(i\rho)}\right|
\label{IdentityTheta1}
\end{eqnarray}
Taking the second derivative of $\ln{\left|\theta_{\alpha,\beta}(z,i\rho)\right|}$
with respect to $z$ at $z=0$ and using the heat equation Eq. (\ref{heat})
we obtain
\begin{eqnarray}
&& \frac{\partial}{\partial\rho}\left({\tt Re} \sum_{n=0}^{\infty}
\sum_{m=1}^{\infty}\left\{\frac{1}{n+\alpha} e^{-2\pi m\big[\,\rho
\left(n+\alpha\right)+i\beta\,\big]}+\frac{1}{n+1-\alpha}
e^{-2\pi m\big[\,\rho\left(n+1-\alpha\right)+i\beta\,\big]}\right\}\right) =
\nonumber\\
&&= -2 \frac{\partial}{\partial\rho}
\ln{\left|\theta_{\alpha,\beta}(i\rho)\right|} +
\frac{1}{2\pi}\left(\frac{\partial}{\partial z}
\ln{\left|\theta_{\alpha,\beta}(i\rho)\right|}\right)^2
\label{IdentityTheta2}
\end{eqnarray}
\section{Kronecker's Double Series}
\label{KroneckerDoubleSeries}

Kronecker's double series can be defined as \cite{Weil}
$${\rm K}_{p}^{\alpha,\beta}(\tau)= -\frac{p!}{(-2\pi i)^p}
\sum_{m,n\in Z \above0pt  (m,n)\neq(0,0)} \frac{e^{-2\pi
i(n\alpha+m\beta)}}{(n+\tau m)^{p}}$$
In this form, however, they cannot be directly applied to our
analysis. We need to cast them in a different form. To this end,
let us separate from the double series a subseries with $m=0$
$${\rm K}_{p}^{\alpha,\beta}(\tau)=-\frac{p!}{(-2\pi i)^p}
\sum_{n\neq 0}\frac{e^{-2\pi in\alpha}}{n^{p}} -\frac{p!}{(-2\pi
i)^p} \sum_{m\neq 0}\sum_{n\in Z} \frac{e^{-2\pi
i(n\alpha+m\beta)}}{(n+\tau m)^{p}}$$
Here the first sum gives nothing but Fourier representation of
Bernoulli polynomials Eq.(\ref{FourierBernoulli}). The second sum
can be rearranged with the help of the identity
$$\frac{p!}{(-2\pi i)^p}\sum_{n\in Z}\frac{e^{-2\pi
in\alpha}}{(z+n)^p}= p\sum_{n=0}^{\infty}(n+\alpha)^{p-1}e^{2\pi i
z(n+\alpha)}$$
which can easily be derived from
Eq.(\ref{FourierGeneratingFunction}) differentiating it $p$ times.
The final result of our resummation of double Kronecker sum is
$${\rm K}_{p}^{\alpha,\beta}(\tau)={\rm B}_{p}^\alpha-p\sum_{m\neq
0}\sum_{n=0}^{\infty} (n+\alpha)^{p-1}~e^{2\pi im
(\tau(n+\alpha)-\beta)}$$
Considering real part of the Kronecker sums with pure imaginary
aspect ratio, $\tau=i\rho$, we can further rearrange this expression
to get summation only over positive $m\geq 1$
\begin{eqnarray}
{\rm B}_{2p}^\alpha-{\tt Re}~\!{\rm
K}_{2p}^{\alpha,\beta}(i\rho)&=&\frac{(2p)!}{(-4\pi^2)^p}~{\tt Re}
\sum_{m=1}^{\infty}\sum_{n\in Z} \frac{e^{-2\pi
i(n\alpha+m\beta)}+ e^{-2\pi i(n(1-\alpha)-m\beta)}}{(n+i\rho
m)^{2p}}\nonumber\\ &=&2p\;{\tt Re}\!\left\{\sum_{m=1}^{\infty}
\sum_{n=0}^{\infty} (n+\alpha)^{2p-1}~e^{-2\pi m
(\rho(n+\alpha)+i\beta)}\right.\nonumber\\&&+\left.\sum_{m=1}^{\infty}
\sum_{n=0}^{\infty} (n+1-\alpha)^{2p-1}~e^{-2\pi m
(\rho(n+1-\alpha)+i\beta)}\right\} \label{IdentityKronecker}
\end{eqnarray}

\section{Asymptotic Expansion of the Digamma Function}
\label{PsiSeries}

Let us start with well known expansion of the digamma function $\psi(N)$
\cite{GradshteinRyzhik}
\begin{eqnarray}
\psi(x)&=&\ln{x}-\frac{1}{2x}-\sum_{p=1}^{\infty}\frac{{\rm B}_{2p}}{2p}
\frac{1}{x^{2p}}
\nonumber\\
&=& \ln{x}-\sum_{p=1}^{\infty}(-1)^p \frac{{\rm B}_{p}}{p}
\frac{1}{x^{p}}
\end{eqnarray}
Plugging in the above expansion $x=N+\alpha$ and expand the resulting factors
$\ln{(1+\alpha/N)}$, $(1+\alpha/N)^{-p}$ in powers of
$N^{-1}$ we obtain
\begin{eqnarray}
\psi(N+\alpha)&=&\ln{N}-\sum_{p=1}^{\infty}(-1)^p
\frac{\alpha^p}{p N^p}-\sum_{p=1}^{\infty}\sum_{k=0}^{\infty}
(-1)^{k+p} {\rm B}_{p} \frac{(p+k-1)!}{k!p!}\frac{\alpha^k}{N^{p+k}}
\nonumber\\
&=& \ln{N}-\sum_{p=1}^{\infty}(-1)^p
\frac{\alpha^p}{p N^p}-\sum_{l=1}^{\infty}\sum_{p=1}^{l}
(-1)^{l} {\rm B}_{p} \frac{(l-1)!}{(l-p)!p!}\frac{\alpha^{l-p}}{N^{l}}
\nonumber\\
&=&\ln{N}- \sum_{l=1}^{\infty}\sum_{p=0}^{l}
(-1)^{l} {\rm B}_{p} \frac{(l-1)!}{(l-p)!p!}\frac{\alpha^{l-p}}{N^{l}}
\end{eqnarray}
Using the relation between Bernoulli polynomials ${\rm B}_{p}^{\alpha}$ and
Bernoulli numbers ${\rm B}_{p}$
\begin{equation}
{\rm B}_{l}^{\alpha}=\sum_{p=0}^{l}{\rm B}_{p}\frac{l!}{(l-p)!p!}
\alpha^{l-p}
\end{equation}
we finally obtain the Eq. (\ref{ExpansionPsi})
\begin{equation}
\psi(N+\alpha)=\ln{N}-\sum_{p=1}^{\infty}(-1)^p
\frac{{\rm B}_{p}^\alpha}{p}\frac{1}{N^p}
\end{equation}

\section{Reduction of Kronecker's Double Series to Theta Functions}
\label{KroneckerToTheta}

Let us consider Laurent expansion of the Weierstrass function
\begin{eqnarray}
\wp(z)&=&\frac{1}{z^2}+\sum_{(n,m)\neq(0,0)}\left[\frac{1}{(z-n-\tau
m)^2}-\frac{1}{(n+\tau m)^2}\right]\nonumber\\
&=&\frac{1}{z^2}+\sum_{p=2}^{\infty}a_{p}(\tau) z^{2p-2}\nonumber
\end{eqnarray}
The coefficients $a_p(\tau)$ of the expansion can all be written
in terms of the elliptic $\theta$-functions with the help of the
recursion relation \cite{Korn}
$$a_p=\frac{3}{(p-3)(2p+1)}~
(a_{2}a_{p-2}+a_{3}a_{p-3}+\ldots+a_{p-2}a_{2})$$
where first terms of the sequence are
\begin{eqnarray}
a_2&=&\textstyle{\frac{\pi^4}{15}}(\theta_2^4\theta_3^4-
\theta_2^4\theta_4^4+\theta_3^4\theta_4^4)\nonumber\\
a_3&=&\textstyle{\frac{\pi^6}{189}}(\theta_2^4+\theta_3^4)
(\theta_4^4-\theta_2^4)(\theta_3^4+\theta_4^4)\nonumber\\
a_4&=&\textstyle{\frac{1}{3}}a_2^2\nonumber\\
a_5&=&\textstyle{\frac{3}{11}}(a_2a_3)\nonumber\\
a_6&=&\textstyle{\frac{1}{39}}(2a_2^3+3a_3^2)\nonumber\\
&\vdots&\nonumber
\end{eqnarray}
Kronecker functions ${\rm K}_{2p}^{0,0}(\tau)$ are related
directly to the coefficients $a_p(\tau)$
$${\rm K}^{0,0}_{2p}(\tau)=
-\frac{(2p)!}{(-4\pi^2)^p}\frac{a_p(\tau)}{(2p-1)}$$
Kronecker functions ${\rm K}_{2p}^{\alpha,\beta}(\tau)$ with
$\alpha$ and $\beta$ taking values $0$ or $1/2$ can in their turn
be related to the function ${\rm K}_{2p}^{0,0}(\tau)$ by means of
simple resummation of Kronecker's double series
\begin{eqnarray}
{\rm K}_{p}^{0,\frac{1}{2}}(\tau)&=&2\,{\rm
K}^{0,0}_{p}(2\tau)-{\rm K}^{0,0}_{p}(\tau)\nonumber\\ {\rm
K}_{p}^{\frac{1}{2},0}(\tau)&=&2^{1-p}\,{\rm
K}^{0,0}_{p}(\tau/2)-{\rm K}^{0,0}_{p}(\tau)\nonumber\\ {\rm
K}_{p}^{\frac{1}{2},\frac{1}{2}}(\tau)&=& (1+2^{2-p})\,{\rm
K}^{0,0}_{p}(\tau)-2^{1-p}\,{\rm K}^{0,0}_{p}(\tau/2)-2\,{\rm
K}^{0,0}_{p}(2\tau)\nonumber
\end{eqnarray}
Thus, Kronecker functions ${\rm K}_{2p}^{\alpha,\beta}(\tau)$ with
$\alpha$ and $\beta$ taking values $0$ or $1/2$ can all be
expressed in terms of the elliptic $\theta$-functions only. For
practical calculations the following identities are also helpful

\begin{center}
\begin{tabular}{rclcrcl}
$2\theta_2^2(2\tau)$&$=$&$\theta_3^2-\theta_4^2$&~~~~~~~~~~~~~
&$\theta_2^2(\tau/2)$&$=$&$2\theta_2\theta_3$\\[0.15cm]
$2\theta_3^2(2\tau)$&$=$&$\theta_3^2+\theta_4^2$&
&$\theta_3^2(\tau/2)$&$=$&$\theta_2^2+\theta_3^2$\\[0.15cm]
$2\theta_4^2(2\tau)$&$=$&$2\theta_3\theta_4$&
&$\theta_4^2(\tau/2)$&$=$&$\theta_3^2-\theta_2^2$
\end{tabular}
\end{center}

From the general formulas above we can easily write down all the
Kronecker functions that have appeared in our asymptotic
expansions
\begin{eqnarray}
{\rm K}_{4}^{0,0}(\tau)&=&
\textstyle{\frac{1}{30}(\theta_2^4\theta_4^4-\theta_2^4\theta_3^4-\theta_3^4\theta_4^4)}\nonumber\\
{\rm
K}_{4}^{0,\frac{1}{2}}(\tau)&=&\textstyle{\frac{1}{30}(\frac{7}{8}\,\theta_2^8-\theta_3^4\theta_4^4)}\nonumber\\
{\rm
K}_{4}^{\frac{1}{2},0}(\tau)&=&\textstyle{\frac{1}{30}(\frac{7}{8}\,\theta_4^8-\theta_2^4\theta_3^4)}\nonumber\\
{\rm
K}_{4}^{\frac{1}{2},\frac{1}{2}}(\tau)&=&\textstyle{\frac{1}{30}(\frac{7}{8}\,\theta_3^8+\theta_2^4\theta_4^4)}\nonumber
\end{eqnarray}
\begin{eqnarray}
 {\rm K}_{6}^{0,0}(\tau)&=&\textstyle{\frac{1}{84}
(\theta_2^4+\theta_3^4)(\theta_4^4-\theta_2^4)(\theta_3^4+\theta_4^4)}\nonumber\\
{\rm K}_{6}^{0,\frac{1}{2}}(\tau)&=&\textstyle{\frac{1}{84}
(\theta_3^4+\theta_4^4)(\frac{31}{16}\,\theta_2^8+\theta_3^4\theta_4^4)}\nonumber\\
{\rm K}_{6}^{\frac{1}{2},0}(\tau)&=&-\textstyle{\frac{1}{84}
(\theta_2^4+\theta_3^4)(\frac{31}{16}\,\theta_4^8+\theta_2^4\theta_3^4)}\nonumber\\
{\rm
K}_{6}^{\frac{1}{2},\frac{1}{2}}(\tau)&=&\textstyle{\frac{1}{84}
(\theta_2^4-\theta_4^4)(\frac{31}{16}\,\theta_3^8-\theta_4^4\theta_2^4)}\nonumber\\
&\vdots&\nonumber
\end{eqnarray}
Note that when $\rho\to\infty$ we have limits $\theta_2\to 0$,
$\theta_4\to 1$, $\theta_3\to 1$ and the Kronecker's function
reduce to the Bernoulli polynomials.

We are grateful to V.B. Priezzhev for critical remarks on the
manuscript. This work was supported by the Russian Foundation for
Basic Research through grant No. 99-01-00882, by the Swiss
National Science Foundation through grant No. 7-SUPJ-62295 and by
the National Science Counsil of the Republic of China (Taiwan)
through grant No. NSC89-2112-M-001-084.

\end{document}